\documentclass[conference]{IEEEtran}
\usepackage{graphicx}
\pdfoutput=1
\begin{document}
\title{Securing Information-Centric Networking without negating Middleboxes}

\author{\IEEEauthorblockN{Nikos Fotiou, George Xylomenos and George C. Polyzos}
\IEEEauthorblockA{Mobile Multimedia Laboratory \& Department of Informatics\\
School of Information Sciences and Technology\\
Athens University of Economics and Business\\
76 Patision, Athens 10434, Greece\\
Email: \{fotiou,xgeorge,polyzos\}@aueb.gr}
}
\maketitle

\begin{abstract}
Information-Centric Networking is a promising networking paradigm that 
overcomes many of the limitations of current networking architectures.
Various research efforts investigate solutions for securing ICN. Nevertheless,
most of these solutions relax security requirements in favor of network 
performance. In particular, they weaken end-user privacy and the architecture's
tolerance to security breaches in order to support middleboxes that offer services
such as caching and content replication. In this paper, we adapt TLS, a widely
used security standard, to an ICN context. We design solutions that allow
session \emph{reuse} and \emph{migration} among multiple stakeholders and
we propose an extension that allows authorized middleboxes to lawfully and transparently intercept
secured communications. 
\end{abstract}

\section{Introduction}
\label{sec:intro}
Information-Centric Networking (ICN) brings the promise of a more secure
and privacy friendly Internet. Indeed, the inherent security properties 
of ICN, including the requirement for explicit signaling of information availability and
demand, are expected to significantly reduce phishing, spamming, and (D)DoS
attacks. Nevertheless, this does not mean that ICN architectures are inherently secure, or that
no additional security measures are required. On the contrary, ICN 
seems to lack a comprehensive security solution. This
can be attributed to two main reasons. Firstly, ICN departs from the traditional
endpoint oriented communication and adopts a content-centric communication
paradigm. This shift in the communication plane impedes the (direct) adoption
of existing security solutions and protocols. Secondly, ICN promotes the \emph{use of middleboxes}
(e.g., caches, content replication points) and \emph{information flow aggregation} (e.g., by using multicast),
which cannot be easily combined with security and privacy enhancing mechanisms.

In this paper we propose a TLS-based security protocol which allows introducing middleboxes in an ICN architecture. In particular,
we propose mechanisms that can be used for transforming TLS from a host oriented protocol
to a content oriented one, allowing different phases of the protocol (concerning the same piece of content) to be executed by different (authenticated) endpoints. 
Moreover, we present an extension to our protocol that allows authorized
middleboxes to \emph{lawfully} and \emph{transparently} intercept secured
transactions. 

The structure of the remainder of this paper is as follows:
In Section~\ref{sec:reference}
we define a generic ICN model which is used as a reference architecture
for our solution. In Section~\ref{sec:overview} we detail the
proposed solution. In Section~\ref{sec:middle} we present an extension to our solution
that enables middleboxes. In Section~\ref{sec:related} we discuss ICN security 
solutions, as well as, the problems middleboxes face when end-to-end
security solutions are used. Finally, we evaluate our solution in Section~\ref{sec:eval},
from both the security and performance perspectives. We provide a summary and present our conclusions in Section~\ref{sec:conclusion}.
  
\section{ICN reference architecture}
\label{sec:reference}
Our solution is not bound to a particular ICN approach. Instead,
in this section we define a generic ICN reference architecture that can be mapped
to most ICN proposals. Our model consists of the following entities: 
\begin{itemize}
\item \emph{Owner}: The entity that creates and owns a \emph{content item}. 
The owner is responsible for assigning \emph{names} to content items.
The role of the owner captures real world entities (e.g., an author, a university, 
a company, a government).
\item \emph{Subscriber}: A network device owned by a real word entity that is 
interested in receiving a content item
\item \emph{Publisher}: A network device that actually hosts a content item.
\end{itemize}
All content items are identified by a \emph{name}. A content name in our model
is composed of two parts: a prefix which is an ICN routable name and is used by
all (inter-)networking functions, and a suffix which can be hidden from the
network. Henceforth, the term \emph{name} will refer to the
\emph{prefix of a content item name}.  

Our model entities interact with each other in the following manner: An owner 
creates a content item and makes available a copy of it to at 
least one publisher. Publishers advertise the names of the content items they host. 
Subscribers \emph{subscribe} to content item names in which they are interested.
A subscription triggers a process (which is out of the scope of our model)
that results in a publisher issuing a \emph{publication} that contains the 
desired item. This publications is \emph{forwarded} by the network to the intended subscriber(s). 
A publication message includes a \emph{forwarding identifier}.
Forwarding identifiers are used by forwarding devices in order to take proper
forwarding decisions. When we refer to a content item name, we will
use the notation ``name'', whereas when we refer to a forwarding identifier,
we will use the notation [name]. In many ICN architectures ``name''==[name],
i.e., the name of a content item is also used as a forwarding identifier. 
 
\subsection{Security model}
Each content owner owns a public-private key pair. The public keys of content
owners are considered known to the subscribers (or, there exists a secure mechanism
for learning them).  Our security model considers two types of publishers: trusted publishers and
regular publishers. A \emph{trusted} publisher has been \emph{authorized} by a content owner
to store certain content items. This authorization can be proven using an x.509
digital certificate signed by the content owner. This certificate includes the content
``names'' for which the publisher is authorized, using the \textit{subject alternative name} 
extension~\cite{Coo2002}. Moreover, this certificate can be used for establishing 
secure connections. A \emph{regular} publisher just happens
to store a content item, but cannot prove it is authorized
to store that item. Nevertheless, in the general case these publishers do
not act maliciously. A regular publisher can be a cache, a content replication
point, and generally any middlebox. In Section~\ref{sec:middle} we present a mechanism with which
a regular publisher can be become authorized for a specific transaction.

\section{Solution Overview}
\label{sec:overview}

The cornerstone of our solution is a handshake protocol that results in the establishment
of a symmetric encryption key and if necessary of an HMAC key,\footnote{If authenticated symmetric encryption is used,
e.g., AES-GCM, then an HMAC algorithm is not required} between a subscriber and a publisher. These
keys are then used for protecting the transmitted content. Our handshake 
protocol is an adaptation of the TLS handshake protocol (section~7.4 of~\cite{Exi2008}).   

Our handshake protocol is composed of four messages: two subscriptions and two publications. In the following 
we consider that both subscriptions are received by the same trusted publisher. In Section~\ref{sec:multi} 
we present solutions that can be used when this is not possible/desirable (i.e., these
solutions allow a subscriber to begin the handshake with a publisher and
finish it with another). The subscriptions that are part of the handshake protocol must not be aggregatable. This can be achieved using various mechanisms,
e.g., by not aggregating -- in general -- subscriptions that do not have the 
same payload, by using a ``special'' flag, or by appending a statistically unique \textit{nonce} to 
the content name included in the subscription. Moreover, the publications that are part of the handshake
protocol should not be cachable, as a cached publication will result in an unsuccessful 
handshake.\footnote{It is the equivalent of ``injecting'' a previously captured server TLS handshake response.}

The protocol starts with the subscriber sending a subscription for the 
desired item. The payload of this subscription message contains the fields of the 
\texttt{Client Hello} TLS message. In order to achieve forward secrecy
the preferred key exchange algorithm must always be (elliptic-curve) 
ephemeral Diffie-Hellman--(EC)DHE. 

The subscription message is routed to a publisher. Assuming that this is a new session (we discuss session reuse
in Section~\ref{sec:multi}), the publisher responds with a publication.
This publication contains the fields of the \texttt{Server Hello},
\texttt{Server Certificate}, \texttt{Server Key Exchange},     
\texttt{Certificate Request}, and \texttt{Server Hello Done} TLS messages.
This messages also includes a signature that can be validated using the publisher's
certificate. 

With the reception of this publication the subscriber verifies that a publisher 
is trusted by validating the received digital signature. If the validation
is successful, the subscriber sends a new 
subscription for the same item. This time the subscription payload
contains the fields of the \texttt{Client Key Exchange},
\texttt{Change Cipher Spec}, and \texttt{Finished} TLS messages. 

When the publisher receives this subscription it should be able to associate it 
with the state generated during the the first step of the handshake. Currently, TLS server implementations 
use the [source IP, source port] pair of the received TLS packets in order
to maintain state. However, not all ICN architectures
have similar packet fields, i.e., a field that contains a subscriber-specific location identifier
that makes possible for a publisher to tell if two subscriptions have originated from the
same subscriber. If no such information exists then two approaches can be considered:
if in the first subscription message a nonce was appended to the content
name (in order to prevent subscription aggregation) then the
same nonce can be used in this subscription; alternatively, the subscriber
can include in the subscription payload the TLS \emph{session identifier}.  
The publisher completes the handshake by responding to this subscription with a
publication that contains the fields of the \texttt{Change Cipher Spec} and \texttt{Finish} TLS 
 messages. 

\subsection{Session migration}
\label{sec:multi}

We now consider the case where content items can be provided by a \emph{group} of
trusted publishers and we discuss how a subscriber that has initiated a
session with a group member can continue it with another.
In order to implement this functionality we use the session ticket TLS
extension~\cite{Sal2008}. With this extension a publisher can encrypt all 
secret information and parameters of a session in a \textit{ticket},
store the key required to decrypt the ticket locally, and transmit the 
ticket to the subscriber; then the subscriber
may use the ticket to request the reuse of previous settings. In order
for all group members to be able to decrypt a ticket they should agree on
the same encryption key. This key can be randomly and periodically generated by a 
\textit{key generator} and proactively published to the group members. 
Twitter for example, uses a similar approach in which a set of key generator 
machines generate and store in all twitter's web servers a fresh ticket
encryption key every 12 hours~\cite{Hoff2013}. It should be noted, however, 
that this approach relaxes the requirement for perfect forward secrecy: a stolen
ticket encryption key can be used for decrypting all sessions that used
tickets encrypted with this key (e.g., in the case of twitter, all user
sessions within a 12 hours window).

\begin{figure*}
\includegraphics[width=0.8\linewidth]{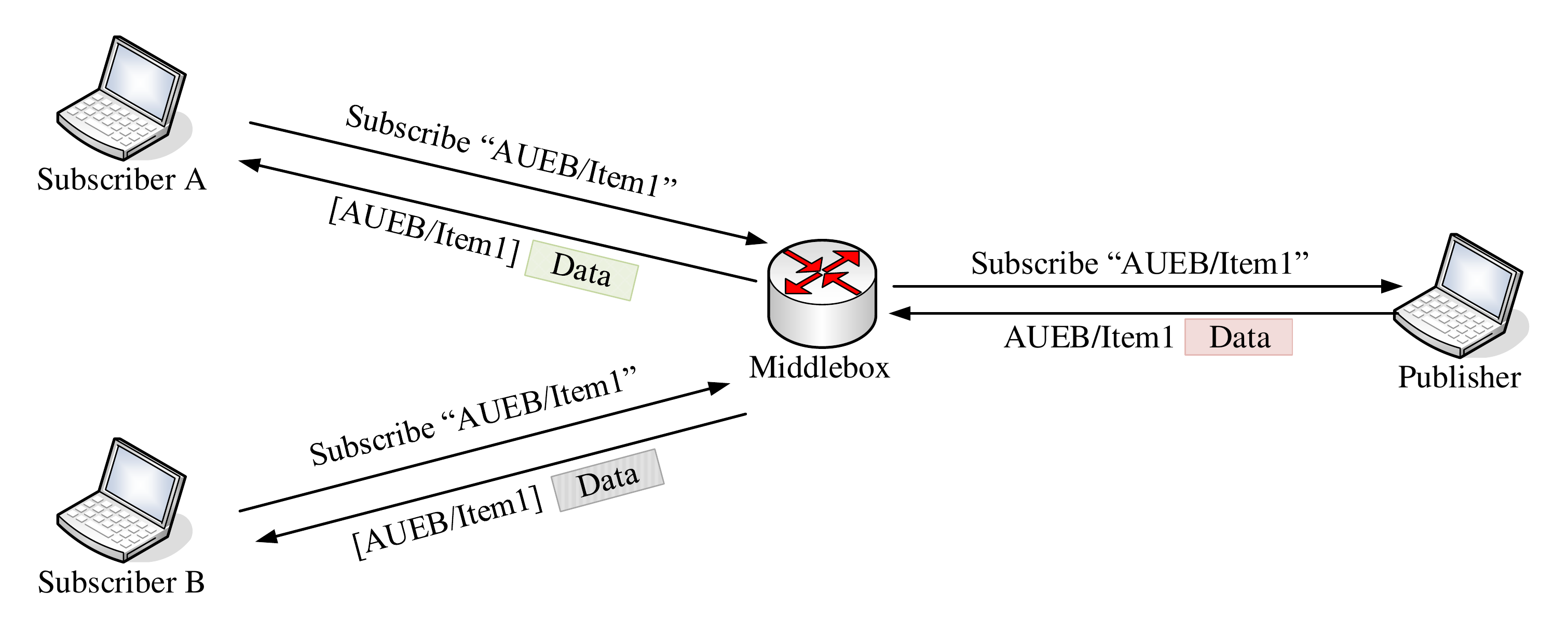}
\caption {A middlebox that lawfully intercepts a secured communication session to
aggregate  subscriptions for the same content item.}
\label{fig:multicast}
\end{figure*}

\section{Enabling middleboxes}
\label{sec:middle}

We now describe an extension  to our solution that allows a regular publisher to
lawfully intercept secured communications. With this extension a regular publisher
is authorized by a trusted publisher to intercept a \emph{single} session. 
That regular publisher (henceforth referred to as the middlebox)
learns only secret information related to the intercepted session, i.e., 
it does not learn any long lasting secret.  
Moreover, this extension requires no modification of the protocol
in the subscriber's side, therefore, subscribers are completely oblivious to
middleboxes. These middleboxes are equipped with a trusted publisher certificate
but not with the secret information associated with this certificate.

With our handshake protocol a subscriber is able to tell if a publisher is trusted by using the 
digital signature included in the first publication of the handshake protocol. This verification is an integral part of the
handshake protocol and a handshake must not be considered successful if this step fails. A middlebox 
is not able to generate this signature, since it 
requires the trusted publisher's secret key. In order to successfully complete the handshake, 
a middlebox should: (i) establish a secured session 
with the trusted publisher, and (ii) send a \textit{signature request} that 
contains the name of the requested content and the data to be signed.
Upon receiving the signature request the trusted publisher first checks 
if the middlebox can handle a subscription for this content name.
If this check is successful, the trusted publisher uses its secret key and 
signs the necessary fields. Finally, it sends the
digital signature back to the middlebox.
As a next step, the middlebox sends a publication back to the subscriber that includes the digital signature.
Now, the handshake can be successfully completed. It should be noted that a 
middlebox does not have to establish a secured session with the trsusted publisher
for every subscriber: an established secured session can be used
for transmitting multiple signature requests. Moreover, step 
(ii) has to take place for every new session,
therefore, a trusted publisher can at any time refuse a signature request,
stopping this way a specific transaction.

Figure~\ref{fig:multicast} illustrates an interesting use case of this extension. In this use case,
two subscribers subscribe to the same content item. A middlebox intercepts 
both these subscriptions and at the same time subscribes to
the desired item. After all subscriptions are completed there is a single data
flow from the trusted publisher to the middlebox; the middlebox decrypts and re-encrypts
this flow and transmits it to the subscribers. An interesting security 
property of this use case is that a third party observing
the encrypted data of these flows is not able to tell whether these flows concern the same 
content or not. 

A notable property of this extension is that trusted publishers can keep track
of item access statistics: everytime a middlebox sends a signature request, it
includes that name of the content item concerned, hence, a trusted publisher
can track the number of times a content item has been requested. 
\subsection{HTTPS-to-ICN gateway}
This extension  can also be applied to legacy TLS. An interesting application 
that can be built using our extension is a middlebox that acts as a
HTTPS-to-ICN gateway. This gateway could allow a legacy HTTPS client to retrieve
a content item from (or through) an ICN network. If plain HTTP were used then 
the approach described in~\cite{Tro2015} could have been used; with this approach the gateway translates 
the web server's domain name into a content item name and issues the appropriate
subscription. When TLS is used the gateway should map the destination IP-port of the first TLS packet 
to a domain name. This is required not only for the domain to content name translation but also
because the gateway should respond to the client with a server certificate.
Fortunately, the Server Name Indication TLS extension~\cite{Eas2011} can be used to solve
this problem. With this extension a TLS client includes in its \texttt{hello}
message the domain name of the server with which it wants to connect.
Another consideration is how the HTTPS client can verify the integrity of the received item. 
This problem is related to the fact that the gateway can manipulate in an undetectable way
a content item. This problem should be solved in the application layer, 
for example the client could retrieve directly from the server the item's
hash and compare it with the hash of the item it received. In any case, no
modification is required to the TLS client.    

\section{Evaluation}
\label{sec:eval}
\subsection{Performance evaluation}
\begin{figure}
\includegraphics[width=0.98\linewidth]{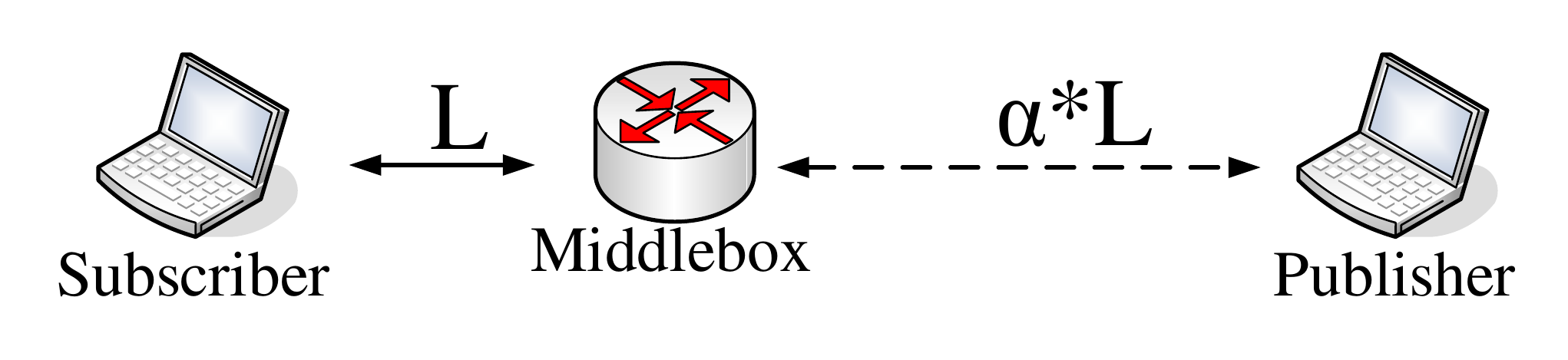}
\caption {Sample evaluation topology.}
\label{fig:eval}
\end{figure}
\begin{figure}
\includegraphics[width=0.98\linewidth]{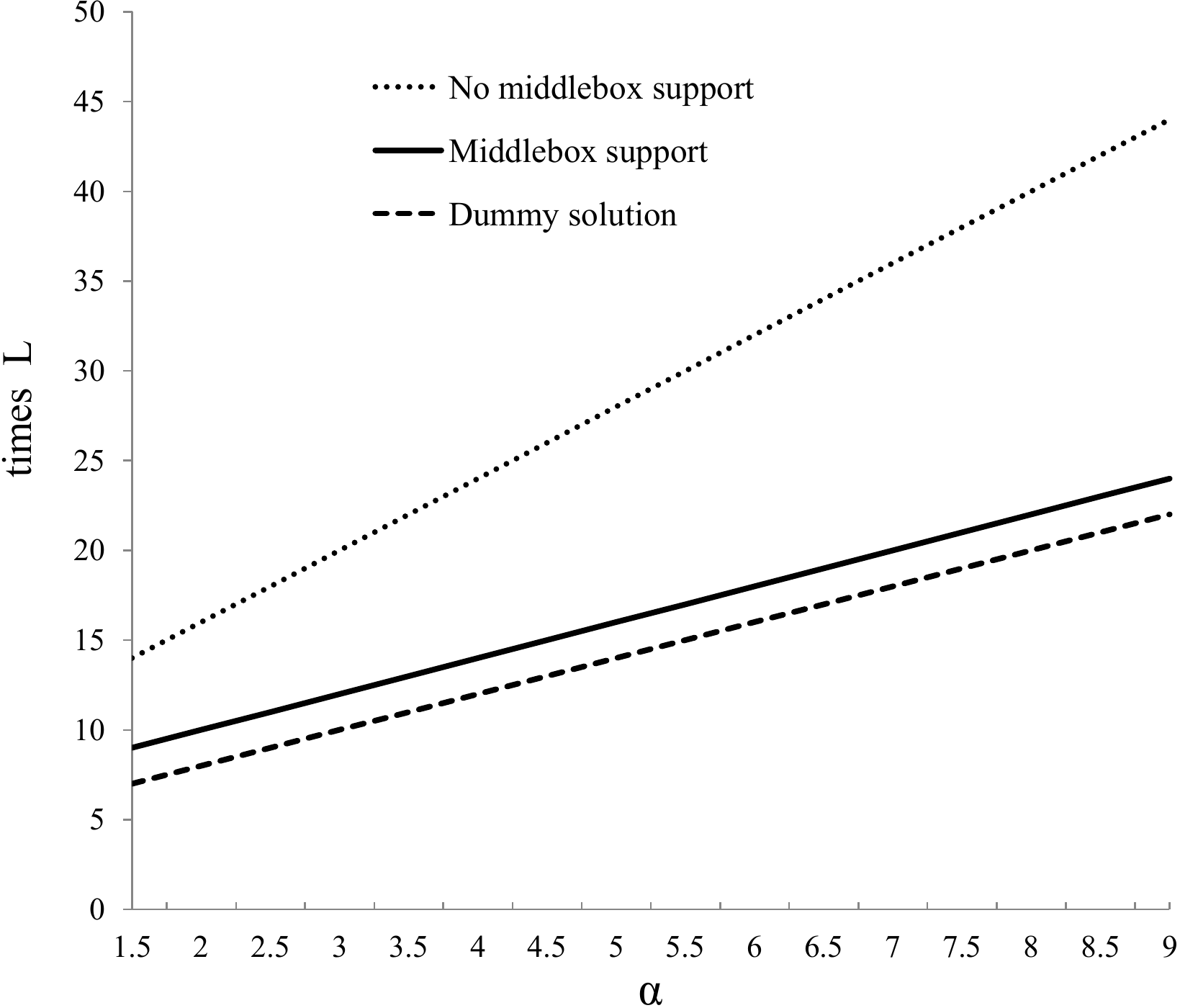}
\caption {Time required for the first packet of the publication to reach the subscriber (measured in multiples of the
latency L) as a function of $\alpha$.}
\label{fig:eval2}
\end{figure}

We now examine the delay that is introduced by our solution. We consider 
the sample topology of Figure~\ref{fig:eval}. In this figure there is a 
subscriber, a middlebox that acts as content replication point, and a
trusted publisher. The subscriber wants to subscribe to an item that is stored
in both the middlebox and the publisher. The latency of the link between the
subscriber and the middlebox is $L$, whereas the latency of the link between
the middlebox and the publisher is $\alpha*L$, where $\alpha$ is a variable. We assume that
the latency of the links is the same for both directions. We compare the time required 
for the subscriber to receive the first packet of the publication when three
security solutions are used: our solution without middlebox support, our
solution with middlebox support, and a dummy solution described in the
following. When the dummy solution is used, the content is encrypted with a 
symmetric encryption key and stored in the middlebox. A subscriber sends simultaneously a subscription message
to the middlebox and a key request message to the publisher. Therefore,
the delay introduced in this case is the time required for the key request
message to reach the publisher plus the time required for the publisher response
to reach the subscriber, i.e., $2*(L + \alpha*L)$.\footnote{In all cases, we consider that the processing time
is negligible.} When our solution is used, we assume that the first packet 
arrives immediately after the last publication of the handshake. Figure~\ref{fig:eval2}
shows the delay introduced by each solution measured as a function of $\alpha$. 
\subsection{Security evaluation}
The security of our solution without middlebox support relies on the security
properties of TLS and the negotiated security algorithms. In particular, our
solution provides the following security properties:
\begin{itemize}
\item \emph{Verification that a publisher is trusted.} With our solution
a subscriber is able to tell is a publisher is authorized to host a particular
content item. In contrast to host-oriented architectures, there is no
notion of publisher identity and hence publisher authentication. 
\item \emph{Content confidentiality and integrity.} With our solution a
symmetric encryption key and (if necessary) an HMAC key are produced. These
keys can be used for securing content confidentiality and integrity.
\item \emph{Unobservability and unlinkability.} A malicious $3^{rd}$ party 
that monitors network flows (or even tampers with them)
is able to tell only the prefix of the content items names and not the full name (unobservability).
Moreover, a malicious $3^{rd}$ party is not able to tell 
whether two distinct network flows concern the same content item, or not (unlinkability).
These two properties are closely related to end-user \emph{privacy}.
\item \emph{Perfect forward secrecy.} With our solution
there is no single secret value 
that can be used to compromise multiple, already completed transactions
\end{itemize}
When middleboxes are
supported, the following security properties hold:
\begin{itemize}
\item \emph{Trusted publishers have control of the process.} Indeed, if a trusted publisher
refuses to accept a signature request the handshake will never be completed. Therefore,
a publisher may even completely refuse to use middleboxes. 
\item \emph{The middlebox learns no trusted publisher-related secret.} The only secret
information that a middlebox learns are the session secrets, negotiated with the 
subscriber.
\end{itemize}

Nevertheless some security precautions should be considered. The trusted publisher has
no control over the negotiated security algorithms, therefore subscriber implementations
should make sure that only secure algorithms are supported. Moreover, a middlebox
is able to modify a content item without being detected, therefore content 
authenticity mechanisms should be in place. 
Another security measure that should be considered when trusted publishers allow middleboxes to intercept secured communication, is to make
sure that no content item that may jeopardize subscriber security is offered via a middlebox.
For example, middleboxes should be prevented from handling subscribers' credentials.
This can be achieved by naming all sensitive content items using a pre-defined name  and
by making sure that trusted publishers will always prevent middleboxes from intercepting
a session that concerns such an item.  
\section{Related work}
\label{sec:related}
\subsection{ICN security solutions}
In general, most ICN related security solutions follow the same pattern: a piece
of content is encrypted with a symmetric key and then this key is encrypted
in a way that only legitimate subscribers can decrypt it. Using the same encryption key
for all subscribers has the advantage that encrypted content can be cached. Nevertheless,
it creates serious security threats. 
In~\cite{Sha2014} a publisher periodically generates a symmetric 
encryption key and encrypts it with the public keys of all legitimate 
subscribers. It then uses this key to encrypt all content items. If 
the private key of a subscriber is compromised then all previous messages 
can be decrypted, therefore this scheme does not provide forward secrecy. 
Moreover, it is possible for a third party to determine if two users 
received the same piece of content, therefore this scheme does provide unlinkability. 
In~\cite{Man2015} a publisher splits a content item in chunks 
and encrypts each chunk using two keys. The second key changes periodically 
in a way that given a key at time $t$, all keys generated at any time $l < t$ can be 
generated. The second key is the same for a ``group of subscribers'', 
therefore it is possible to distinguish if two subscribers that belong to 
the same group have received the same content item. Moreover, this key 
is delivered to authorized subscribers by following a key 
exchange protocol during which the key is encrypted using a session key. 
The session key is generated by the subscriber and it is transmitted to 
the publisher encrypted using the publisher's public key. Therefore, if the
publisher's private key is compromised all previous communication can be decrypted. Wood and Uzun~\cite{Woo2014}
encrypt content items with a symmetric key and then encrypt the symmetric keys
with a public key encryption scheme that facilitates proxy re-encryption.
Again, it is possible for an attacker to link two information flows.
Moreover, if the symmetric key is compromised all previous transactions
can be decrypted. The work in~\cite{Zha2011} uses Identity-Based Encryption
in order to encrypt a symmetric encryption key that has been used for encrypting
a piece of content. As in all previous related work, there is no unlinkability.
Moreover, if the private key of a subscriber is compromised, all previous
communication can be decrypted.

\subsection{Middlebox support}

As HTTPS traffic increases, middleboxes become less effective. In order 
to mitigate this problem, the IETF HTTPBis Working Group has drafted a 
proposal for ``explicit trusted'' proxies~\cite{Lor2014}. Explicit 
trusted proxies are authorized by end-users to fetch secured content 
on behalf of them. Nevertheless, this solution has many shortcomings: 
end-users are involved in the proxy selection process, HTTP clients 
have to be modified and web servers have no control over the process. 
Sherry et al.~\cite{Sher2015} follow another approach. In their solution,
a server includes in every TLS packet some ``attributes'' of the
(encrypted) content item e.g., its type, possible age restrictions etc. These tokens are encrypted
in a form that only authorized middleboxes can inspect. Nevertheless, these
middleboxes have very limited control over the encrypted content.

With our solution, middleboxes have full access to the content, therefore they 
are able to cache it or manipulate it. Middlebox support is completely transparent to 
subscribers, i.e., a subscriber is not able to tell whether it interacts with
a middlebox or a publisher. Moreover, a publisher has full control of the process, i.e., 
at any time it may prevent a middlebox for intercepting secured communication. Finally,
a middlebox learns no publisher-related secrets.   
\section{Conclusion}
\label{sec:conclusion}
In this paper we proposed a TLS-based security solution for ICN. In particular,
we proposed modifications to TLS so that it can be effectively used in a content-centric environment.
We chose to modify TLS, instead of creating a clean slate solution, since TLS is an industry standard and its security properties are continuously examined and tested by many researchers around the world.
The main challenges which we came across were:
how to allow the network to transparently migrate a process from one publisher to another,
as well as, how to enable middlebox support without modifying the subscriber-side part of the protocol. 

\section*{Acknowledgment}
This research was supported by the EU funded H2020 ICT project POINT, under contract 643990.



\begin{thebibliography}{10}
\providecommand{\url}[1]{#1}
\csname url@samestyle\endcsname
\providecommand{\newblock}{\relax}
\providecommand{\bibinfo}[2]{#2}
\providecommand{\BIBentrySTDinterwordspacing}{\spaceskip=0pt\relax}
\providecommand{\BIBentryALTinterwordstretchfactor}{4}
\providecommand{\BIBentryALTinterwordspacing}{\spaceskip=\fontdimen2\font plus
\BIBentryALTinterwordstretchfactor\fontdimen3\font minus
  \fontdimen4\font\relax}
\providecommand{\BIBforeignlanguage}[2]{{%
\expandafter\ifx\csname l@#1\endcsname\relax
\typeout{** WARNING: IEEEtran.bst: No hyphenation pattern has been}%
\typeout{** loaded for the language `#1'. Using the pattern for}%
\typeout{** the default language instead.}%
\else
\language=\csname l@#1\endcsname
\fi
#2}}
\providecommand{\BIBdecl}{\relax}
\BIBdecl

\bibitem{Coo2002}
D.~Cooper \emph{et~al.}, ``Internet {X}.509 {P}ublic {K}ey {I}nfrastructure
  {C}ertificate and {C}ertificate {R}evocation {L}ist {P}rofile,'' IETF, RFC
  5280, 2008.

\bibitem{Exi2008}
T.~Dierks and E.~Rescorla, ``{The Transport Layer Security (TLS) Protocol
  Version 1.2},'' \emph{RFC 5246}, August 2008.

\bibitem{Sal2008}
J.~Salowey, H.~Zhou, P.~Eronen, and H.~Tschofenig, ``{T}ransport {L}ayer
  {S}ecurity ({TLS}) session resumption without server-side state,'' IETF, RFC
  5077, 2008.

\bibitem{Hoff2013}
J.~Hoffman-Andrews, ``Forward secrecy at {T}witter,'' \emph{Twitter Engineering
  {B}log}, November 2013,
  https://blog.twitter.com/2013/forward-secrecy-at-twitter.

\bibitem{Tro2015}
D.~Trossen \emph{et~al.}, ``{IP Over ICN - The Better IP}? {A}n unusual take on
  {Information-Centric Networking},'' in \emph{Proceedings of the EUCNC
  conference}, 2015.

\bibitem{Eas2011}
D.~Eastlake, ``{Transport Layer Security (TLS) Extensions: Extension
  Definitions},'' \emph{RFC 6066}, January 2011.

\bibitem{Sha2014}
W.~Shang, Q.~Ding, A.~Marianantoni, J.~Burke, and L.~Zhang, ``Securing building
  management systems using named data networking,'' \emph{Network, IEEE},
  vol.~28, no.~3, pp. 50--56, May 2014.

\bibitem{Man2015}
M.~Mangili, F.~Martignon, and S.~Paraboschi, ``A cache-aware mechanism to
  enforce confidentiality, trackability and access policy evolution in
  content-centric networks,'' \emph{Computer Networks}, vol.~76, no.~0, pp. 126
  -- 145, 2015.

\bibitem{Woo2014}
C.~Wood and E.~Uzun, ``Flexible end-to-end content security in {CCN},'' in
  \emph{Consumer Communications and Networking Conference (CCNC), 2014 IEEE
  11th}, Jan 2014, pp. 858--865.

\bibitem{Zha2011}
X.~Zhang, K.~Chang, H.~Xiong, Y.~Wen, G.~Shi, and G.~Wang, ``Towards name-based
  trust and security for content-centric network,'' in \emph{Network Protocols
  (ICNP), 2011 19th IEEE International Conference on}, Oct 2011, pp. 1--6.

\bibitem{Lor2014}
{S. Loreto et al.}, ``{Explicit Trusted Proxy in HTTP/2.0},''
  \emph{Internet-Draft}, August 2014.

\bibitem{Sher2015}
J.~Sherry, C.~Lan, R.~A. Popa, and S.~Ratnasamy, ``Blindbox: Deep packet
  inspection over encrypted traffic,'' Cryptology ePrint Archive, Report
  2015/264, 2015, http://eprint.iacr.org/.

\end{thebibliography}
\end{document}